\documentclass[reprint,
superscriptaddress,
 amsmath,amssymb,
aps,
floatfix,
]{revtex4-2}

\usepackage{graphicx}
\usepackage{dcolumn}
\usepackage{bm}
\usepackage{float}
\usepackage{graphicx}

\usepackage{hyperref}

\begin{document}

\preprint{APS/123-QED}

\title{A tuneable wavelength reference for chip-scale laser cooling}

\author{S. Dyer}
 \affiliation{SUPA and Department of Physics, University of Strathclyde, G4 0NG, Glasgow, United Kingdom}
 
 \author{K. Gallacher}%
\affiliation{James Watt School of Engineering, University of Glasgow, Rankine Building, Oakfield Avenue, Glasgow, G12 8LT, U.K.}

\author{U. Hawley}%
\affiliation{James Watt School of Engineering, University of Glasgow, Rankine Building, Oakfield Avenue, Glasgow, G12 8LT, U.K.}
 
 \author{A. Bregazzi}
 \affiliation{SUPA and Department of Physics, University of Strathclyde, G4 0NG, Glasgow, United Kingdom}

\author{P. F. Griffin}%
 \affiliation{SUPA and Department of Physics, University of Strathclyde, G4 0NG, Glasgow, United Kingdom}
 
 \author{A. S. Arnold}%
 \affiliation{SUPA and Department of Physics, University of Strathclyde, G4 0NG, Glasgow, United Kingdom}

\author{D. J. Paul}%
\affiliation{James Watt School of Engineering, University of Glasgow, Rankine Building, Oakfield Avenue, Glasgow, G12 8LT, U.K.}

\author{E. Riis}%
 \affiliation{SUPA and Department of Physics, University of Strathclyde, G4 0NG, Glasgow, United Kingdom}

\author{J. P. McGilligan}%
 \affiliation{SUPA and Department of Physics, University of Strathclyde, G4 0NG, Glasgow, United Kingdom}

\date{\today}

\begin{abstract}
We demonstrate a tuneable, chip-scale wavelength reference to greatly reduce the complexity and volume of cold-atom sensors. A 1$\,$mm optical path length micro-fabricated cell provides an atomic wavelength reference, with dynamic frequency control enabled by Zeeman shifting the atomic transition through the magnetic field generated by the printed circuit board (PCB) coils. The dynamic range of the laser frequency stabilization system is evaluated and used in conjunction with an improved generation of chip-scale cold atom platforms that traps 4 million $^{87}$Rb atoms. The scalability and component consolidation provide a key step forward in the miniaturization of cold atom sensors.
\end{abstract}

\maketitle

\section{\label{sec:intro}Introduction}

The miniaturization of cold atom platforms for remote sensing and portability has tantalising prospects for next-generation quantum technology \cite{Takamoto2020,bidel,gravitybrum,mcgilliganreview,Bao2012}. Owing to the long interrogation times achievable with laser cooling, the development of on-chip cold-atom platforms could improve the stability of current thermal atomic sensors to the benefit of applications in navigation, geological surveying, communication and precision timing, driving towards fully integrated atomic systems. 

Recent research has demonstrated drastic improvements in the scalability of critical components at the heart of atomic sensors. While initially focused on the scalability of thermal atom packages, the micro-fabrication of core components revolutionized the scalability in a plethora of instruments, including atomic clocks \cite{newman2021highperformance,Knappe2004,boudotclock}, magnetometers \cite{microfabcoilkitching,mitchellcoils,Hunter:22}, and wavelength references \cite{Hummon:18,Kitching2018}. At the heart of many of these devices is a micro-electro-mechanical systems (MEMS) vapor cell. Typically formed from a glass-silicon-glass wafer stack, the on-chip thermal vapor system offers a simplicity and design versatility that complements its application in precision spectroscopy \cite{moreland,P_tremand_2012, Chutani2015,karlencell}.

The miniaturization of thermal atom packages has also driven forward the scalability of cold-atom sensors. Recent studies have demonstrated cold-atom sensor scalability through the coupling of photonic integrated circuits \cite{McGehee_2021,Chauhan:19,Chauhan:22}, meta-surfaces \cite{Yulaev:21,metasurface}, diffractive optics \cite{Nshii2013,Henderson:20,McGilligan2017}, planar coil systems \cite{chen2021planar} and micro-fabricated alkali sources \cite{Kang:19,mcgilliganAIB}. More recently, MEMS vapor cells capable of sustaining ultra-high-vacuum (UHV) have been shown in conjunction with micro-fabricated grating chips as an on-chip platform for cold-atoms \cite{McGilligan2020}. Parallel avenues of research have aimed to address the miniaturization and simplicity of cold atom systems through the development of plug-and-play interfaces, making use of conventional off-the-shelf optics \cite{plugplay,3Dprint}. Many of these cold-atom systems, however, have focused on individual component development, such that a fully integrated cold-atom platform has remained elusive. 

In this Article, we demonstrate a tuneable wavelength reference at 780$\,$nm that complements chip-scale laser cooling of Rb atoms. A single optical package enables laser frequency tuning and stabilization in a micro-fabricated vapor cell, sandwiched between a printed-circuit-board (PCB) coil pair. The dynamic range of the chip-scale package is evaluated as a viable replacement to standard larger components used for real-time frequency control, such as acousto-optical modulators. The wavelength reference is coupled with a chip-scale cold-atom platform to demonstrate device simplicity and compatibility for laser-cooling applications. Finally, the combined apparatus demonstrates laser cooled atom numbers $>10^6$ in thick silicon MEMS vacuum cells. The union of such components provides a key step in the scalability and mass production of cold-atom systems.

\section{\label{sec:exp}Experimental set-up}
\begin{figure*}[t!]
\centering
\includegraphics[width=0.9\textwidth]{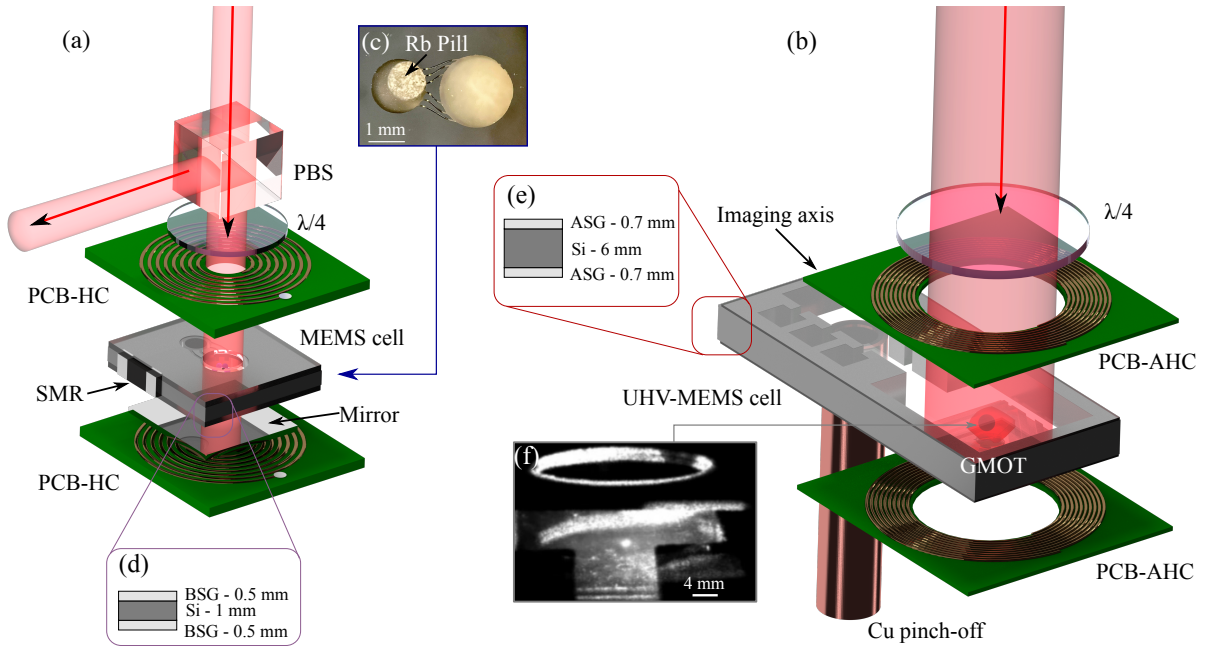}
\caption{\label{satspec} A schematic diagram overview of the chip-scale wavelength and cooling platforms. (a): A MEMS cell based wavelength reference using saturated absorption spectroscopy. PCB-HC: Printed-circuit-board Helmholtz coils. PBS: Polarizing beam splitter. SMR: Surface mounted resistor. (b): MEMS cell-GMOT incorporated laser cooling platform with planar coils. GMOT: Grating magneto-optical trap. PCB-AHC: Printed-circuit-board anti-Helmholtz coils.(c): Microscope image of the MEMS cell in the wavelength reference package. (d): Layer structure for the wavelength reference MEMS cell stack. BSG: Borosilicate glass. Si: Silicon. (e): Layer structure for the UHV-MEMS cell stack. ASG: Aluminosilicate glass. (f): Example image of the cold-atom sample formed from the combined chip-scale platforms.}
\end{figure*}

The experimental apparatus used for the wavelength reference and chip-scale laser cooling platform are shown in Fig.~\ref{satspec} (a) and (b) respectively. The light source is provided from a butterfly packaged volume-Bragg reflector (VBR) laser centred at 780$\,$nm. The light from the VBR is incident through a 5$\,$mm polarizing beam splitter (PBS), with one arm coupled into our chip-scale wavelength reference. 

The laser wavelength reference is provided by saturated absorption spectroscopy through a 2$\,$mm thick MEMS cell, shown at the centre of Fig.~\ref{satspec} (a). The MEMS spectroscopy cell, shown in Fig.~\ref{satspec} (c), is composed of a 1$\,$mm thick silicon frame anodically bonded at both interfaces to 0.5$\,$mm thick borosilicate glass wafers, as highlighted in Fig.~\ref{satspec} (d). The silicon frame is deep reactive-ion etched using an anisotropic Bosch process \cite{Middlemiss2016} to form a two-chamber cell connected via 650$\,\mu$m wide, non-line-of-sight channels. The first chamber houses a micro-pill based alkali vapor source (SAES Rb/AMAX/Pill1-0.6). The second chamber acts as the spectroscopy region with a 2$\,$mm diameter aperture. The cell is mechanically diced to dimensions of $10\,$mm$\times10\,$mm$\times2\,$mm and a surface-mounted resistor is adhered to the side of the silicon frame to provide heating. Prior to the implementation of the MEMS cell in the apparatus, the pill is laser activated with 1$\,$W of 1070$\,$nm laser light focused onto the pill surface for 10$\,$s \cite{GriffLaser}.

The MEMS cell is sandwiched between a pair of PCB Helmholtz coils. Each 1$\,$mm thick PCB is double-side printed and electrically connected via a circuit through-hole to increase the number of turns over the given board volume. The coil is formed of 35$\,$$\mu$m thick copper layer wires with 170$\,\mu$m width deposited on the surfaces of the board to form a total of 14 turns, with a mean radius of 3.3$\,$mm ($R_{min}=2.2\,$mm, $R_{max}=4.5\,$mm). The centre of the coil has a 1.5$\,$mm radius aperture that is aligned for optical interrogation of the MEMS cell through the spectroscopy region. The coils and cell are held firmly in location within a 3-D printed mount, to ensure a robust alignment. The laser is frequency modulated at 250$\,$kHz via current dithering. The output from the laser is passed through a PBS, circularly polarized with a $\lambda/4$ wave-plate before the MEMS cell and reflected by a dielectric reflector, mounted to the back of the MEMS cell. The output transmission from the spectroscopy system is directed through the PBS to a photodiode, where the absorption signal is passed to external electronics for demodulation and the resulting error signal is used to electronically feed-back to the VBR current for frequency stabilization.

The remaining light is passed through an electro-optical-modulator (EOM) driven at 6.5$\,$GHz to generate 5$\,\%$ side-bands for hyperfine re-pumping in the laser cooling process. The light is then fibre coupled and expanded in free-space to the cold-atom system. The light is circularly polarized with a $\lambda/4$ wave-plate before being incident through the UHV-MEMS cell. The UHV-MEMS cell, shown in Fig.~\ref{satspec} (b), is composed of a 6~mm thick silicon frame, fabricated with water-jet processing to achieve the deep silicon cut \cite{dyer}. The silicon wafer is anodically bonded at the upper and lower surfaces to 700$\,$$\mu$m thick aluminosilicate glass wafers, as highlighted in Fig.~\ref{satspec} (e). The upper glass wafer is connected to an ion pump and resistively-heated alkali dispenser through a copper pinch-off tube adhered over a mechanically drilled glass through-hole. The UHV-MEMS cell outer dimensions are 70$\,$mm$\,\times34\,$mm$\,\times7.4\,$mm with a 20$\,$mm$\times20\,$mm$\times6\,$mm laser cooling region, which accommodates the surface area of the grating chip. The 3-segment grating chip is fabricated with a 1100$\,$nm period over the 2$\,$cm$\times\,$2$\,$cm surface area of the chip. At the center of the grating, a 2$\,$mm diameter hole has been laser cut for optical access and imaging \cite{Bregazzi}. The grating chip is placed immediately below the UHV-MEMS cell lower window to ensure a maximum optical overlap volume exists within the UHV-MEMS cell vacuum volume. A 30$\,$mm$\times\,$8$\,$mm channel in the cell from the pinch-off region leading to the cooling region was designed to aid fluorescence imaging of the cold-atoms by allowing a low angular displacement between the cell and imaging system. The low incident angle for imaging avoids diffracted orders from the grating chip and greatly reduces the surface scatter from the cell walls, glass and grating. An example of this imaging axis is shown in Fig.~\ref{satspec} (f). The UHV-MEMS cell and grating are sandwiched by a PCB anti-Helmholtz pair (PCB-AHC). These PCB coils are of the same thickness and wire properties as the previously described Helmholtz PCBs. Each PCB-AHC, however, is formed of 30 turns, 15 on each side, with a mean radius of 12.5$\,$mm ($R_{min}\,=\,10.2\,$mm, $R_{max}=14.9\,$mm).

To enable miniaturization of the cold-atom optical system, the bulky optical and electronic system required for the acoustic optical modulator (AOM) was replaced with the tuneable Zeeman lock \cite{GaetanoZeeman}. The laser is locked directly to the $F=2\rightarrow F'=3$ cooling transition, and a frequency offset is provided from the PCB Helmholtz coils, with the magnetic field orientation illustrated in Fig.~\ref{Bshift} (a). The magnetic field is simulated from a numerical solution to the Biot-Savart law, for parameters meeting that of the PCB. The offset mechanism uses a bias magnetic field along the axis of the beam, which induces a shift to the Zeeman sub-levels. Circularly polarized light is used to optically pump the atoms into the stretched states. As such, the magnetically sensitive sub-Doppler features can be exploited to tune the laser lock frequency while avoiding the optical losses typical of a double passed AOM setup \cite{AOMtune}. To determine the versatility of the frequency offset, a beat-note was measured against an external-cavity diode-laser (ECDL), which was stabilized to the $F=2\rightarrow F'=2,3$ cross-over transition of $^{87}$Rb using saturated absorption spectroscopy on a separate reference cell. The beat-note signal was obtained by overlapping light from each laser onto a fast photodiode and the output frequency was tracked on a frequency counter.

\section{\label{sec:res}Results}
\begin{figure}[t!]
\centering
\includegraphics[width=0.45\textwidth]{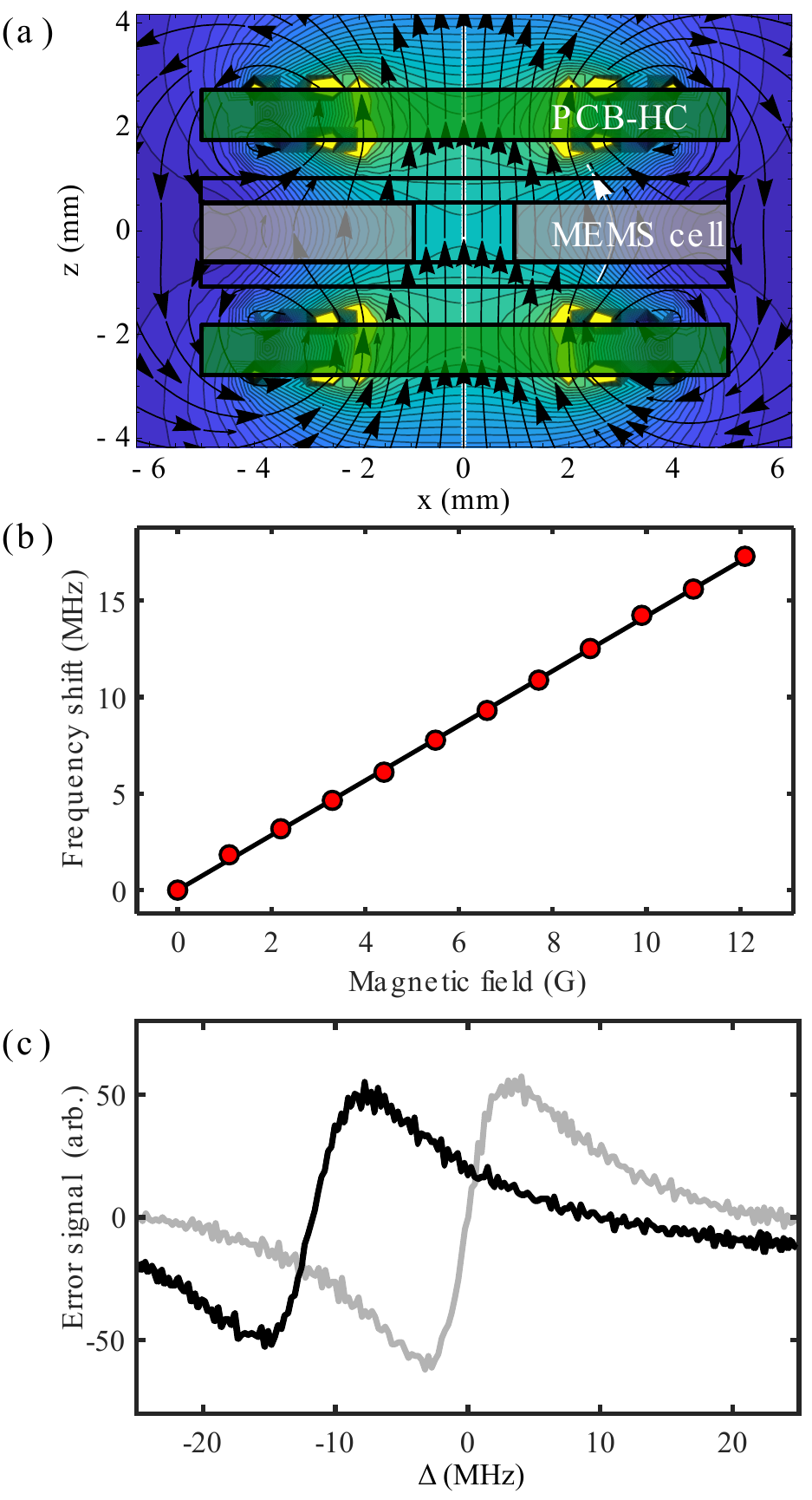}
\caption{\label{Bshift} (a): Printed-circuit-board generated Helmholtz magnetic field orientation through the 2$\,$mm thick MEMS cell. (b): Measured frequency shift as a function of the applied magnetic field, as calculated from the coil properties. (c): The Zeeman shifted error signal at 12$\,$MHz detuning (black) and on resonance (gray) with 16 averages.}
\end{figure}

The induced frequency shift for weak magnetic fields was found to be $1.42\pm0.12\,$MHz/G in agreement with the expected shift for the $F=2,m_F=2 \rightarrow F'=3,m_F=3$ transition, indicating stretched-state optical pumping. For stronger magnetic fields, the error signal is deformed due to lineshape broadening from magnetic field inhomogeneities, placing a limit on the frequency tuning range of $\approx\pm$30$\,$MHz. However, this range is sufficient to achieve a red-detuned lock for the magneto-optical trap, where the trapped atom number is typically optimum in Rb for a detuning between $\Delta=-\Gamma\rightarrow-2\Gamma$, where $\Gamma$ is the natural linewidth of the excited state ($\Gamma_{\textrm{D$_2$}}=2\pi\times6.1$$\,$MHz) \cite{Steck}. Over this detuning range, the PCB-HC coil pair require $<$$\,$0.7$\,$W of electrical power. The error signal from the MEMS cell, shown for $\Delta=12$$\,$MHz in Fig.~\ref{Bshift} (c) with a black line, has no significant amplitude or linewidth degradation compared to the resonant signal, shown in gray. The detuning range here could also prove beneficial to the frequency detunings typical of optical molasses.

\begin{figure*}[t!]
\centering
\includegraphics[width=0.95\textwidth]{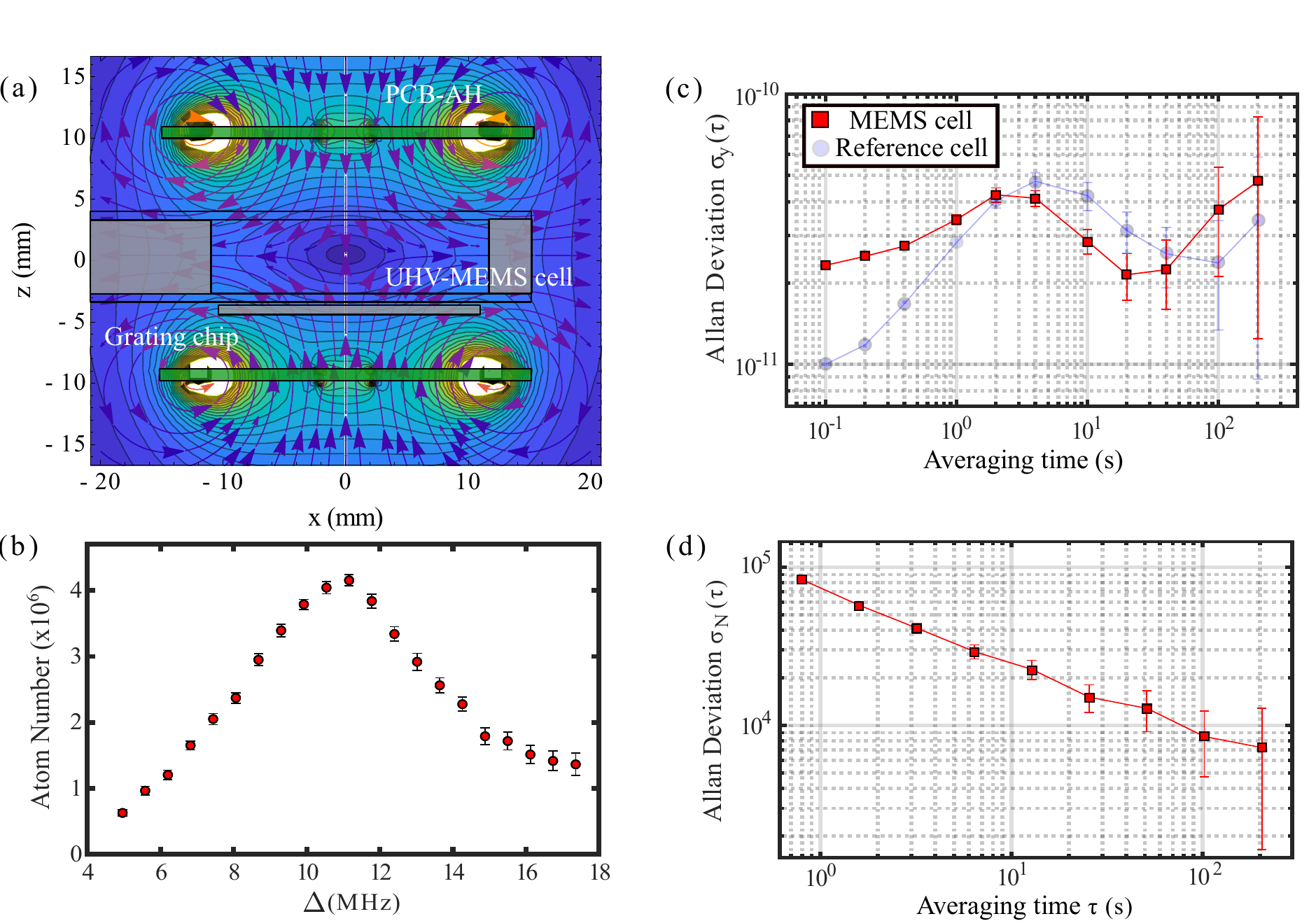}
\caption{\label{Atom_stability} (a): Printed-circuit-board generated anti-Helmholtz magnetic field orientation through the 6$\,$mm thick UHV-MEMS cell. (b): Measured atom number as a function of laser lock frequency, provided by the MEMS wavelength reference. (c): Overlapping Allan deviation of the beat-note signal when the VBR laser is stabilised to the tuneable MEMS wavelength reference and a 7$\,$cm reference cell, shown in red squares and blue circles respectively. (d): Simultaneously measured atom number stability from the MEMS-GMOT platform.}
\end{figure*}

To demonstrate the applicability of the wavelength reference to cold-atom sensors, the VBR was locked to the cooling transition, and coupled into the chip-scale cold-atom platform. To aid further miniaturization, traditionally large wire-wound anti-Helmholtz coils have been replaced in favour of a PCB solution. Each double-sided PCB anti-Helmholtz coil used in the cold-atom system has a resistance of 10$\,$$\Omega$ and draws 1$\,$A of current to achieve 20$\,$G/cm, with the field orientation relative to the other system components shown in Fig.~\ref{Atom_stability} (a). 

The increased vacuum cell volume of the UHV-MEMS cell, compared to previous work, has been made possible from the recently demonstrated water-jet cutting of silicon wafers \cite{dyer}. The improved cell volume now enables $\approx 93$\,$\%$ of the GMOT optical overlap volume to coincide within the MEMS cell \cite{Bregazzi}. Due to this, an optimum atom number of 4$\times$10$^6$ atoms is obtained in the chip-scale platform for a detuning of 11$\,$MHz, as demonstrated in Fig.~\ref{Atom_stability} (b), agreeing well with previous GMOT work \cite{Nshii2013}. It was observed that with an incident intensity of $\approx$$\,$13$\,$mW/cm$^2$, the atom number would reach a maximum at 20$\,$G/cm.

Finally, the frequency fluctuations of the chip-scale wavelength reference were temporally measured, with the impact on the measured atom number from the cold-atom platform simultaneously tracked. During this measurement the detuning was set to 12$\,$MHz, which produced an atom number of $\approx\,3\times10^6$. The relative stability of the wavelength reference is illustrated in Fig.~\ref{Atom_stability} (c). The observed relative frequency stability is on the order of 10$^{-11}$ for all integration windows. While this stability does not meet the performance of state-of-the-art on-chip wavelength reference \cite{Hummon:18}, it is suitable for keeping the laser drift well below the 6$\,$MHz linewidth of the cooling transition to satisfy the miniaturization of the cold-atom apparatus. Further to this, when the tuneable wavelength reference is replaced with a 7$\,$cm reference cell, a similar stability was observed, as demonstrated in Fig.~\ref{Atom_stability} (c), indicating that the stability is not limited by the MEMS cell package. The simultaneously measured atom number stability, shown in Fig.~\ref{Atom_stability} (d), integrates down with a slope of $\tau^{-1/2}$, demonstrating that the frequency fluctuations of the wavelength reference have no contribution to the stability of the trapped atom number.

The amalgamation of these on-chip technologies has successfully demonstrated a significant scalability in cold-atom technology, while offering a simplified optical system comprised of planar components.

\section{\label{sec:con}Conclusion}
We have demonstrated an on-chip solution for wavelength referencing in cold-atom systems. The simple architecture and mass producible components are suitable for implementation in a wide range of quantum technologies, with a detuning range of $\pm$30$\,$MHz demonstrated in constant use, and increased values possible for short time intervals. The amalgamation with the 6$\,$mm thick silicon based UHV-MEMS cell directly demonstrates the applicability of the tuneable wavelength reference as a suitable replacement for traditionally bulky cold-atom optical systems.

\begin{acknowledgments}
The authors would like to thank R.\ Boudot for useful conversations. The authors acknowledge funding from Defence Security and Technology Laboratory, Engineering and Physical Sciences Research Council (EP/T001046/1), and Defence and Security Accelerator. The authors would like to acknowledge support from the INMAQS collaboration (EP/W026929/1). J. P. M gratefully acknowledges funding from a Royal Academy of Engineering Research Fellowship and D. J. P. from a Royal Academy of Engineering Research Chair in Emerging Technologies (CiET2021$\backslash$123). A. B. was supported by a PhD studentship from the Defence Science and Technology Laboratory (Dstl).
\end{acknowledgments}

\appendix

\bibliography{library}

\providecommand{\noopsort}[1]{}\providecommand{\singleletter}[1]{#1}%
\begin{thebibliography}{38}%
\makeatletter
\providecommand \@ifxundefined [1]{%
 \@ifx{#1\undefined}
}%
\providecommand \@ifnum [1]{%
 \ifnum #1\expandafter \@firstoftwo
 \else \expandafter \@secondoftwo
 \fi
}%
\providecommand \@ifx [1]{%
 \ifx #1\expandafter \@firstoftwo
 \else \expandafter \@secondoftwo
 \fi
}%
\providecommand \natexlab [1]{#1}%
\providecommand \enquote  [1]{``#1''}%
\providecommand \bibnamefont  [1]{#1}%
\providecommand \bibfnamefont [1]{#1}%
\providecommand \citenamefont [1]{#1}%
\providecommand \href@noop [0]{\@secondoftwo}%
\providecommand \href [0]{\begingroup \@sanitize@url \@href}%
\providecommand \@href[1]{\@@startlink{#1}\@@href}%
\providecommand \@@href[1]{\endgroup#1\@@endlink}%
\providecommand \@sanitize@url [0]{\catcode `\\12\catcode `\$12\catcode
  `\&12\catcode `\#12\catcode `\^12\catcode `\_12\catcode `\%12\relax}%
\providecommand \@@startlink[1]{}%
\providecommand \@@endlink[0]{}%
\providecommand \url  [0]{\begingroup\@sanitize@url \@url }%
\providecommand \@url [1]{\endgroup\@href {#1}{\urlprefix }}%
\providecommand \urlprefix  [0]{URL }%
\providecommand \Eprint [0]{\href }%
\providecommand \doibase [0]{https://doi.org/}%
\providecommand \selectlanguage [0]{\@gobble}%
\providecommand \bibinfo  [0]{\@secondoftwo}%
\providecommand \bibfield  [0]{\@secondoftwo}%
\providecommand \translation [1]{[#1]}%
\providecommand \BibitemOpen [0]{}%
\providecommand \bibitemStop [0]{}%
\providecommand \bibitemNoStop [0]{.\EOS\space}%
\providecommand \EOS [0]{\spacefactor3000\relax}%
\providecommand \BibitemShut  [1]{\csname bibitem#1\endcsname}%
\let\auto@bib@innerbib\@empty
\bibitem [{\citenamefont {Takamoto}\ \emph {et~al.}(2020)\citenamefont
  {Takamoto}, \citenamefont {Ushijima}, \citenamefont {Ohmae}, \citenamefont
  {Yahagi}, \citenamefont {Kokado}, \citenamefont {Shinkai},\ and\
  \citenamefont {Katori}}]{Takamoto2020}%
  \BibitemOpen
  \bibfield  {author} {\bibinfo {author} {\bibfnamefont {M.}~\bibnamefont
  {Takamoto}}, \bibinfo {author} {\bibfnamefont {I.}~\bibnamefont {Ushijima}},
  \bibinfo {author} {\bibfnamefont {N.}~\bibnamefont {Ohmae}}, \bibinfo
  {author} {\bibfnamefont {T.}~\bibnamefont {Yahagi}}, \bibinfo {author}
  {\bibfnamefont {K.}~\bibnamefont {Kokado}}, \bibinfo {author} {\bibfnamefont
  {H.}~\bibnamefont {Shinkai}},\ and\ \bibinfo {author} {\bibfnamefont
  {H.}~\bibnamefont {Katori}},\ }\bibfield  {title} {\bibinfo {title} {Test of
  general relativity by a pair of transportable optical lattice clocks},\
  }\href {https://doi.org/10.1038/s41566-020-0619-8} {\bibfield  {journal}
  {\bibinfo  {journal} {Nature Photonics}\ }\textbf {\bibinfo {volume} {14}},\
  \bibinfo {pages} {411} (\bibinfo {year} {2020})}\BibitemShut {NoStop}%
\bibitem [{\citenamefont {Bidel}\ \emph {et~al.}(2013)\citenamefont {Bidel},
  \citenamefont {Carraz}, \citenamefont {Charrière}, \citenamefont {Cadoret},
  \citenamefont {Zahzam},\ and\ \citenamefont {Bresson}}]{bidel}%
  \BibitemOpen
  \bibfield  {author} {\bibinfo {author} {\bibfnamefont {Y.}~\bibnamefont
  {Bidel}}, \bibinfo {author} {\bibfnamefont {O.}~\bibnamefont {Carraz}},
  \bibinfo {author} {\bibfnamefont {R.}~\bibnamefont {Charrière}}, \bibinfo
  {author} {\bibfnamefont {M.}~\bibnamefont {Cadoret}}, \bibinfo {author}
  {\bibfnamefont {N.}~\bibnamefont {Zahzam}},\ and\ \bibinfo {author}
  {\bibfnamefont {A.}~\bibnamefont {Bresson}},\ }\bibfield  {title} {\bibinfo
  {title} {Compact cold atom gravimeter for field applications},\ }\href
  {https://doi.org/10.1063/1.4801756} {\bibfield  {journal} {\bibinfo
  {journal} {Applied Physics Letters}\ }\textbf {\bibinfo {volume} {102}},\
  \bibinfo {pages} {144107} (\bibinfo {year} {2013})}\BibitemShut {NoStop}%
\bibitem [{\citenamefont {Stray}\ \emph {et~al.}(2022)\citenamefont {Stray},
  \citenamefont {Lamb}, \citenamefont {Kaushik}, \citenamefont {Vovrosh},
  \citenamefont {Rodgers}, \citenamefont {Winch}, \citenamefont {Hayati},
  \citenamefont {Boddice}, \citenamefont {Stabrawa}, \citenamefont {Niggebaum},
  \citenamefont {Langlois}, \citenamefont {Lien}, \citenamefont {Lellouch},
  \citenamefont {Roshanmanesh}, \citenamefont {Ridley}, \citenamefont
  {de~Villiers}, \citenamefont {Brown}, \citenamefont {Cross}, \citenamefont
  {Tuckwell}, \citenamefont {Faramarzi}, \citenamefont {Metje}, \citenamefont
  {Bongs},\ and\ \citenamefont {Holynski}}]{gravitybrum}%
  \BibitemOpen
  \bibfield  {author} {\bibinfo {author} {\bibfnamefont {B.}~\bibnamefont
  {Stray}}, \bibinfo {author} {\bibfnamefont {A.}~\bibnamefont {Lamb}},
  \bibinfo {author} {\bibfnamefont {A.}~\bibnamefont {Kaushik}}, \bibinfo
  {author} {\bibfnamefont {J.}~\bibnamefont {Vovrosh}}, \bibinfo {author}
  {\bibfnamefont {A.}~\bibnamefont {Rodgers}}, \bibinfo {author} {\bibfnamefont
  {J.}~\bibnamefont {Winch}}, \bibinfo {author} {\bibfnamefont
  {F.}~\bibnamefont {Hayati}}, \bibinfo {author} {\bibfnamefont
  {D.}~\bibnamefont {Boddice}}, \bibinfo {author} {\bibfnamefont
  {A.}~\bibnamefont {Stabrawa}}, \bibinfo {author} {\bibfnamefont
  {A.}~\bibnamefont {Niggebaum}}, \bibinfo {author} {\bibfnamefont
  {M.}~\bibnamefont {Langlois}}, \bibinfo {author} {\bibfnamefont {Y.-H.}\
  \bibnamefont {Lien}}, \bibinfo {author} {\bibfnamefont {S.}~\bibnamefont
  {Lellouch}}, \bibinfo {author} {\bibfnamefont {S.}~\bibnamefont
  {Roshanmanesh}}, \bibinfo {author} {\bibfnamefont {K.}~\bibnamefont
  {Ridley}}, \bibinfo {author} {\bibfnamefont {G.}~\bibnamefont {de~Villiers}},
  \bibinfo {author} {\bibfnamefont {G.}~\bibnamefont {Brown}}, \bibinfo
  {author} {\bibfnamefont {T.}~\bibnamefont {Cross}}, \bibinfo {author}
  {\bibfnamefont {G.}~\bibnamefont {Tuckwell}}, \bibinfo {author}
  {\bibfnamefont {A.}~\bibnamefont {Faramarzi}}, \bibinfo {author}
  {\bibfnamefont {N.}~\bibnamefont {Metje}}, \bibinfo {author} {\bibfnamefont
  {K.}~\bibnamefont {Bongs}},\ and\ \bibinfo {author} {\bibfnamefont
  {M.}~\bibnamefont {Holynski}},\ }\bibfield  {title} {\bibinfo {title}
  {Quantum sensing for gravity cartography},\ }\href
  {https://doi.org/10.1038/s41586-021-04315-3} {\bibfield  {journal} {\bibinfo
  {journal} {Nature}\ }\textbf {\bibinfo {volume} {602}},\ \bibinfo {pages}
  {590} (\bibinfo {year} {2022})}\BibitemShut {NoStop}%
\bibitem [{\citenamefont {McGilligan}\ \emph {et~al.}(2022)\citenamefont
  {McGilligan}, \citenamefont {Gallacher}, \citenamefont {Griffin},
  \citenamefont {Paul}, \citenamefont {Arnold},\ and\ \citenamefont
  {Riis}}]{mcgilliganreview}%
  \BibitemOpen
  \bibfield  {author} {\bibinfo {author} {\bibfnamefont {J.~P.}\ \bibnamefont
  {McGilligan}}, \bibinfo {author} {\bibfnamefont {K.}~\bibnamefont
  {Gallacher}}, \bibinfo {author} {\bibfnamefont {P.~F.}\ \bibnamefont
  {Griffin}}, \bibinfo {author} {\bibfnamefont {D.~J.}\ \bibnamefont {Paul}},
  \bibinfo {author} {\bibfnamefont {A.~S.}\ \bibnamefont {Arnold}},\ and\
  \bibinfo {author} {\bibfnamefont {E.}~\bibnamefont {Riis}},\ }\bibfield
  {title} {\bibinfo {title} {Micro-fabricated components for cold atom
  sensors},\ }\href {https://doi.org/10.1063/5.0101628} {\bibfield  {journal}
  {\bibinfo  {journal} {Review of Scientific Instruments}\ }\textbf {\bibinfo
  {volume} {93}},\ \bibinfo {pages} {091101} (\bibinfo {year}
  {2022})}\BibitemShut {NoStop}%
\bibitem [{\citenamefont {Bao}\ \emph {et~al.}(2012)\citenamefont {Bao},
  \citenamefont {Reingruber}, \citenamefont {Dietrich}, \citenamefont {Rui},
  \citenamefont {D{\"{u}}ck}, \citenamefont {Strassel}, \citenamefont {Li},
  \citenamefont {Liu}, \citenamefont {Zhao},\ and\ \citenamefont
  {Pan}}]{Bao2012}%
  \BibitemOpen
  \bibfield  {author} {\bibinfo {author} {\bibfnamefont {X.~H.}\ \bibnamefont
  {Bao}}, \bibinfo {author} {\bibfnamefont {A.}~\bibnamefont {Reingruber}},
  \bibinfo {author} {\bibfnamefont {P.}~\bibnamefont {Dietrich}}, \bibinfo
  {author} {\bibfnamefont {J.}~\bibnamefont {Rui}}, \bibinfo {author}
  {\bibfnamefont {A.}~\bibnamefont {D{\"{u}}ck}}, \bibinfo {author}
  {\bibfnamefont {T.}~\bibnamefont {Strassel}}, \bibinfo {author}
  {\bibfnamefont {L.}~\bibnamefont {Li}}, \bibinfo {author} {\bibfnamefont
  {N.~L.}\ \bibnamefont {Liu}}, \bibinfo {author} {\bibfnamefont
  {B.}~\bibnamefont {Zhao}},\ and\ \bibinfo {author} {\bibfnamefont {J.~W.}\
  \bibnamefont {Pan}},\ }\bibfield  {title} {\bibinfo {title} {{Efficient and
  long-lived quantum memory with cold atoms inside a ring cavity}},\ }\href
  {https://doi.org/10.1038/nphys2324} {\bibfield  {journal} {\bibinfo
  {journal} {Nat. Phys.}\ }\textbf {\bibinfo {volume} {8}},\ \bibinfo {pages}
  {517} (\bibinfo {year} {2012})},\ \Eprint {https://arxiv.org/abs/1207.2894}
  {arXiv:1207.2894} \BibitemShut {NoStop}%
\bibitem [{\citenamefont {Newman}\ \emph {et~al.}(2021)\citenamefont {Newman},
  \citenamefont {Maurice}, \citenamefont {Fredrick}, \citenamefont {Fortier},
  \citenamefont {Leopardi}, \citenamefont {Hollberg}, \citenamefont {Diddams},
  \citenamefont {Kitching},\ and\ \citenamefont
  {Hummon}}]{newman2021highperformance}%
  \BibitemOpen
  \bibfield  {author} {\bibinfo {author} {\bibfnamefont {Z.~L.}\ \bibnamefont
  {Newman}}, \bibinfo {author} {\bibfnamefont {V.}~\bibnamefont {Maurice}},
  \bibinfo {author} {\bibfnamefont {C.}~\bibnamefont {Fredrick}}, \bibinfo
  {author} {\bibfnamefont {T.}~\bibnamefont {Fortier}}, \bibinfo {author}
  {\bibfnamefont {H.}~\bibnamefont {Leopardi}}, \bibinfo {author}
  {\bibfnamefont {L.}~\bibnamefont {Hollberg}}, \bibinfo {author}
  {\bibfnamefont {S.~A.}\ \bibnamefont {Diddams}}, \bibinfo {author}
  {\bibfnamefont {J.}~\bibnamefont {Kitching}},\ and\ \bibinfo {author}
  {\bibfnamefont {M.~T.}\ \bibnamefont {Hummon}},\ }\bibfield  {title}
  {\bibinfo {title} {High-performance, compact optical standard},\ }\href
  {https://doi.org/10.1364/OL.435603} {\bibfield  {journal} {\bibinfo
  {journal} {Opt. Lett.}\ }\textbf {\bibinfo {volume} {46}},\ \bibinfo {pages}
  {4702} (\bibinfo {year} {2021})}\BibitemShut {NoStop}%
\bibitem [{\citenamefont {Knappe}\ \emph {et~al.}(2004)\citenamefont {Knappe},
  \citenamefont {Shah}, \citenamefont {Schwindt}, \citenamefont {Hollberg},
  \citenamefont {Kitching}, \citenamefont {Liew},\ and\ \citenamefont
  {Moreland}}]{Knappe2004}%
  \BibitemOpen
  \bibfield  {author} {\bibinfo {author} {\bibfnamefont {S.}~\bibnamefont
  {Knappe}}, \bibinfo {author} {\bibfnamefont {V.}~\bibnamefont {Shah}},
  \bibinfo {author} {\bibfnamefont {P.~D.~D.}\ \bibnamefont {Schwindt}},
  \bibinfo {author} {\bibfnamefont {L.}~\bibnamefont {Hollberg}}, \bibinfo
  {author} {\bibfnamefont {J.}~\bibnamefont {Kitching}}, \bibinfo {author}
  {\bibfnamefont {L.-A.}\ \bibnamefont {Liew}},\ and\ \bibinfo {author}
  {\bibfnamefont {J.}~\bibnamefont {Moreland}},\ }\bibfield  {title} {\bibinfo
  {title} {A microfabricated atomic clock},\ }\href
  {https://doi.org/10.1063/1.1787942} {\bibfield  {journal} {\bibinfo
  {journal} {Applied Physics Letters}\ }\textbf {\bibinfo {volume} {85}},\
  \bibinfo {pages} {1460} (\bibinfo {year} {2004})}\BibitemShut {NoStop}%
\bibitem [{\citenamefont {Hasegawa}\ \emph {et~al.}(2011)\citenamefont
  {Hasegawa}, \citenamefont {Chutani}, \citenamefont {Gorecki}, \citenamefont
  {Boudot}, \citenamefont {Dziuban}, \citenamefont {Giordano}, \citenamefont
  {Clatot},\ and\ \citenamefont {Mauri}}]{boudotclock}%
  \BibitemOpen
  \bibfield  {author} {\bibinfo {author} {\bibfnamefont {M.}~\bibnamefont
  {Hasegawa}}, \bibinfo {author} {\bibfnamefont {R.}~\bibnamefont {Chutani}},
  \bibinfo {author} {\bibfnamefont {C.}~\bibnamefont {Gorecki}}, \bibinfo
  {author} {\bibfnamefont {R.}~\bibnamefont {Boudot}}, \bibinfo {author}
  {\bibfnamefont {P.}~\bibnamefont {Dziuban}}, \bibinfo {author} {\bibfnamefont
  {V.}~\bibnamefont {Giordano}}, \bibinfo {author} {\bibfnamefont
  {S.}~\bibnamefont {Clatot}},\ and\ \bibinfo {author} {\bibfnamefont
  {L.}~\bibnamefont {Mauri}},\ }\bibfield  {title} {\bibinfo {title}
  {Microfabrication of cesium vapor cells with buffer gas for mems atomic
  clocks},\ }\href {https://doi.org/https://doi.org/10.1016/j.sna.2011.02.039}
  {\bibfield  {journal} {\bibinfo  {journal} {Sensors and Actuators A:
  Physical}\ }\textbf {\bibinfo {volume} {167}},\ \bibinfo {pages} {594}
  (\bibinfo {year} {2011})},\ \bibinfo {note} {solid-State Sensors, Actuators
  and Microsystems Workshop}\BibitemShut {NoStop}%
\bibitem [{\citenamefont {Schwindt}\ \emph {et~al.}(2007)\citenamefont
  {Schwindt}, \citenamefont {Lindseth}, \citenamefont {Knappe}, \citenamefont
  {Shah}, \citenamefont {Kitching},\ and\ \citenamefont
  {Liew}}]{microfabcoilkitching}%
  \BibitemOpen
  \bibfield  {author} {\bibinfo {author} {\bibfnamefont {P.~D.~D.}\
  \bibnamefont {Schwindt}}, \bibinfo {author} {\bibfnamefont {B.}~\bibnamefont
  {Lindseth}}, \bibinfo {author} {\bibfnamefont {S.}~\bibnamefont {Knappe}},
  \bibinfo {author} {\bibfnamefont {V.}~\bibnamefont {Shah}}, \bibinfo {author}
  {\bibfnamefont {J.}~\bibnamefont {Kitching}},\ and\ \bibinfo {author}
  {\bibfnamefont {L.-A.}\ \bibnamefont {Liew}},\ }\bibfield  {title} {\bibinfo
  {title} {Chip-scale atomic magnetometer with improved sensitivity by use of
  the mx technique},\ }\href {https://doi.org/10.1063/1.2709532} {\bibfield
  {journal} {\bibinfo  {journal} {Applied Physics Letters}\ }\textbf {\bibinfo
  {volume} {90}},\ \bibinfo {pages} {081102} (\bibinfo {year}
  {2007})}\BibitemShut {NoStop}%
\bibitem [{\citenamefont {Tayler}\ \emph {et~al.}(2022)\citenamefont {Tayler},
  \citenamefont {Mouloudakis}, \citenamefont {Zetter}, \citenamefont {Hunter},
  \citenamefont {Lucivero}, \citenamefont {Bodenstedt}, \citenamefont
  {Parkkonen},\ and\ \citenamefont {Mitchell}}]{mitchellcoils}%
  \BibitemOpen
  \bibfield  {author} {\bibinfo {author} {\bibfnamefont {M.~C.~D.}\
  \bibnamefont {Tayler}}, \bibinfo {author} {\bibfnamefont {K.}~\bibnamefont
  {Mouloudakis}}, \bibinfo {author} {\bibfnamefont {R.}~\bibnamefont {Zetter}},
  \bibinfo {author} {\bibfnamefont {D.}~\bibnamefont {Hunter}}, \bibinfo
  {author} {\bibfnamefont {V.~G.}\ \bibnamefont {Lucivero}}, \bibinfo {author}
  {\bibfnamefont {S.}~\bibnamefont {Bodenstedt}}, \bibinfo {author}
  {\bibfnamefont {L.}~\bibnamefont {Parkkonen}},\ and\ \bibinfo {author}
  {\bibfnamefont {M.~W.}\ \bibnamefont {Mitchell}},\ }\bibfield  {title}
  {\bibinfo {title} {Miniature biplanar coils for alkali-metal-vapor
  magnetometry},\ }\href {https://doi.org/10.1103/PhysRevApplied.18.014036}
  {\bibfield  {journal} {\bibinfo  {journal} {Phys. Rev. Applied}\ }\textbf
  {\bibinfo {volume} {18}},\ \bibinfo {pages} {014036} (\bibinfo {year}
  {2022})}\BibitemShut {NoStop}%
\bibitem [{\citenamefont {Hunter}\ \emph {et~al.}(2022)\citenamefont {Hunter},
  \citenamefont {Dyer},\ and\ \citenamefont {Riis}}]{Hunter:22}%
  \BibitemOpen
  \bibfield  {author} {\bibinfo {author} {\bibfnamefont {D.}~\bibnamefont
  {Hunter}}, \bibinfo {author} {\bibfnamefont {T.~E.}\ \bibnamefont {Dyer}},\
  and\ \bibinfo {author} {\bibfnamefont {E.}~\bibnamefont {Riis}},\ }\bibfield
  {title} {\bibinfo {title} {Accurate optically pumped magnetometer based on
  ramsey-style interrogation},\ }\href {https://doi.org/10.1364/OL.449180}
  {\bibfield  {journal} {\bibinfo  {journal} {Opt. Lett.}\ }\textbf {\bibinfo
  {volume} {47}},\ \bibinfo {pages} {1230} (\bibinfo {year}
  {2022})}\BibitemShut {NoStop}%
\bibitem [{\citenamefont {Hummon}\ \emph {et~al.}(2018)\citenamefont {Hummon},
  \citenamefont {Kang}, \citenamefont {Bopp}, \citenamefont {Li}, \citenamefont
  {Westly}, \citenamefont {Kim}, \citenamefont {Fredrick}, \citenamefont
  {Diddams}, \citenamefont {Srinivasan}, \citenamefont {Aksyuk},\ and\
  \citenamefont {Kitching}}]{Hummon:18}%
  \BibitemOpen
  \bibfield  {author} {\bibinfo {author} {\bibfnamefont {M.~T.}\ \bibnamefont
  {Hummon}}, \bibinfo {author} {\bibfnamefont {S.}~\bibnamefont {Kang}},
  \bibinfo {author} {\bibfnamefont {D.~G.}\ \bibnamefont {Bopp}}, \bibinfo
  {author} {\bibfnamefont {Q.}~\bibnamefont {Li}}, \bibinfo {author}
  {\bibfnamefont {D.~A.}\ \bibnamefont {Westly}}, \bibinfo {author}
  {\bibfnamefont {S.}~\bibnamefont {Kim}}, \bibinfo {author} {\bibfnamefont
  {C.}~\bibnamefont {Fredrick}}, \bibinfo {author} {\bibfnamefont {S.~A.}\
  \bibnamefont {Diddams}}, \bibinfo {author} {\bibfnamefont {K.}~\bibnamefont
  {Srinivasan}}, \bibinfo {author} {\bibfnamefont {V.}~\bibnamefont {Aksyuk}},\
  and\ \bibinfo {author} {\bibfnamefont {J.~E.}\ \bibnamefont {Kitching}},\
  }\bibfield  {title} {\bibinfo {title} {Photonic chip for laser stabilization
  to an atomic vapor with 10$^{11}$ instability},\ }\href
  {https://doi.org/10.1364/OPTICA.5.000443} {\bibfield  {journal} {\bibinfo
  {journal} {Optica}\ }\textbf {\bibinfo {volume} {5}},\ \bibinfo {pages} {443}
  (\bibinfo {year} {2018})}\BibitemShut {NoStop}%
\bibitem [{\citenamefont {Kitching}(2018)}]{Kitching2018}%
  \BibitemOpen
  \bibfield  {author} {\bibinfo {author} {\bibfnamefont {J.}~\bibnamefont
  {Kitching}},\ }\bibfield  {title} {\bibinfo {title} {Chip-scale atomic
  devices},\ }\href {https://doi.org/10.1063/1.5026238} {\bibfield  {journal}
  {\bibinfo  {journal} {Applied Physics Reviews}\ }\textbf {\bibinfo {volume}
  {5}},\ \bibinfo {pages} {031302} (\bibinfo {year} {2018})}\BibitemShut
  {NoStop}%
\bibitem [{\citenamefont {Liew}\ \emph {et~al.}(2004)\citenamefont {Liew},
  \citenamefont {Knappe}, \citenamefont {Moreland}, \citenamefont {Robinson},
  \citenamefont {Hollberg},\ and\ \citenamefont {Kitching}}]{moreland}%
  \BibitemOpen
  \bibfield  {author} {\bibinfo {author} {\bibfnamefont {L.-A.}\ \bibnamefont
  {Liew}}, \bibinfo {author} {\bibfnamefont {S.}~\bibnamefont {Knappe}},
  \bibinfo {author} {\bibfnamefont {J.}~\bibnamefont {Moreland}}, \bibinfo
  {author} {\bibfnamefont {H.}~\bibnamefont {Robinson}}, \bibinfo {author}
  {\bibfnamefont {L.}~\bibnamefont {Hollberg}},\ and\ \bibinfo {author}
  {\bibfnamefont {J.}~\bibnamefont {Kitching}},\ }\bibfield  {title} {\bibinfo
  {title} {Microfabricated alkali atom vapor cells},\ }\href
  {https://doi.org/10.1063/1.1691490} {\bibfield  {journal} {\bibinfo
  {journal} {Applied Physics Letters}\ }\textbf {\bibinfo {volume} {84}},\
  \bibinfo {pages} {2694} (\bibinfo {year} {2004})}\BibitemShut {NoStop}%
\bibitem [{\citenamefont {P{\'{e}}tremand}\ \emph {et~al.}(2012)\citenamefont
  {P{\'{e}}tremand}, \citenamefont {Affolderbach}, \citenamefont {Straessle},
  \citenamefont {Pellaton}, \citenamefont {Briand}, \citenamefont {Mileti},\
  and\ \citenamefont {de~Rooij}}]{P_tremand_2012}%
  \BibitemOpen
  \bibfield  {author} {\bibinfo {author} {\bibfnamefont {Y.}~\bibnamefont
  {P{\'{e}}tremand}}, \bibinfo {author} {\bibfnamefont {C.}~\bibnamefont
  {Affolderbach}}, \bibinfo {author} {\bibfnamefont {R.}~\bibnamefont
  {Straessle}}, \bibinfo {author} {\bibfnamefont {M.}~\bibnamefont {Pellaton}},
  \bibinfo {author} {\bibfnamefont {D.}~\bibnamefont {Briand}}, \bibinfo
  {author} {\bibfnamefont {G.}~\bibnamefont {Mileti}},\ and\ \bibinfo {author}
  {\bibfnamefont {N.~F.}\ \bibnamefont {de~Rooij}},\ }\bibfield  {title}
  {\bibinfo {title} {Microfabricated rubidium vapour cell with a thick glass
  core for small-scale atomic clock applications},\ }\href
  {https://doi.org/10.1088/0960-1317/22/2/025013} {\bibfield  {journal}
  {\bibinfo  {journal} {Journal of Micromechanics and Microengineering}\
  }\textbf {\bibinfo {volume} {22}},\ \bibinfo {pages} {025013} (\bibinfo
  {year} {2012})}\BibitemShut {NoStop}%
\bibitem [{\citenamefont {Chutani}\ \emph {et~al.}(2015)\citenamefont
  {Chutani}, \citenamefont {Maurice}, \citenamefont {Passilly}, \citenamefont
  {Gorecki}, \citenamefont {Boudot}, \citenamefont {Abdel~Hafiz}, \citenamefont
  {Abb{\'e}}, \citenamefont {Galliou}, \citenamefont {Rauch},\ and\
  \citenamefont {de~Clercq}}]{Chutani2015}%
  \BibitemOpen
  \bibfield  {author} {\bibinfo {author} {\bibfnamefont {R.}~\bibnamefont
  {Chutani}}, \bibinfo {author} {\bibfnamefont {V.}~\bibnamefont {Maurice}},
  \bibinfo {author} {\bibfnamefont {N.}~\bibnamefont {Passilly}}, \bibinfo
  {author} {\bibfnamefont {C.}~\bibnamefont {Gorecki}}, \bibinfo {author}
  {\bibfnamefont {R.}~\bibnamefont {Boudot}}, \bibinfo {author} {\bibfnamefont
  {M.}~\bibnamefont {Abdel~Hafiz}}, \bibinfo {author} {\bibfnamefont
  {P.}~\bibnamefont {Abb{\'e}}}, \bibinfo {author} {\bibfnamefont
  {S.}~\bibnamefont {Galliou}}, \bibinfo {author} {\bibfnamefont {J.-Y.}\
  \bibnamefont {Rauch}},\ and\ \bibinfo {author} {\bibfnamefont
  {E.}~\bibnamefont {de~Clercq}},\ }\bibfield  {title} {\bibinfo {title} {Laser
  light routing in an elongated micromachined vapor cell with diffraction
  gratings for atomic clock applications},\ }\href
  {https://doi.org/10.1038/srep14001} {\bibfield  {journal} {\bibinfo
  {journal} {Scientific Reports}\ }\textbf {\bibinfo {volume} {5}},\ \bibinfo
  {pages} {14001} (\bibinfo {year} {2015})}\BibitemShut {NoStop}%
\bibitem [{\citenamefont {Karlen}\ \emph {et~al.}(2020)\citenamefont {Karlen},
  \citenamefont {Haesler}, \citenamefont {Overstolz}, \citenamefont
  {Bergonzi},\ and\ \citenamefont {Lecomte}}]{karlencell}%
  \BibitemOpen
  \bibfield  {author} {\bibinfo {author} {\bibfnamefont {S.}~\bibnamefont
  {Karlen}}, \bibinfo {author} {\bibfnamefont {J.}~\bibnamefont {Haesler}},
  \bibinfo {author} {\bibfnamefont {T.}~\bibnamefont {Overstolz}}, \bibinfo
  {author} {\bibfnamefont {G.}~\bibnamefont {Bergonzi}},\ and\ \bibinfo
  {author} {\bibfnamefont {S.}~\bibnamefont {Lecomte}},\ }\bibfield  {title}
  {\bibinfo {title} {Sealing of mems atomic vapor cells using cu-cu
  thermocompression bonding},\ }\href
  {https://doi.org/10.1109/JMEMS.2019.2949349} {\bibfield  {journal} {\bibinfo
  {journal} {Journal of Microelectromechanical Systems}\ }\textbf {\bibinfo
  {volume} {29}},\ \bibinfo {pages} {95} (\bibinfo {year} {2020})}\BibitemShut
  {NoStop}%
\bibitem [{\citenamefont {McGehee}\ \emph {et~al.}(2021)\citenamefont
  {McGehee}, \citenamefont {Zhu}, \citenamefont {Barker}, \citenamefont
  {Westly}, \citenamefont {Yulaev}, \citenamefont {Klimov}, \citenamefont
  {Agrawal}, \citenamefont {Eckel}, \citenamefont {Aksyuk},\ and\ \citenamefont
  {McClelland}}]{McGehee_2021}%
  \BibitemOpen
  \bibfield  {author} {\bibinfo {author} {\bibfnamefont {W.~R.}\ \bibnamefont
  {McGehee}}, \bibinfo {author} {\bibfnamefont {W.}~\bibnamefont {Zhu}},
  \bibinfo {author} {\bibfnamefont {D.~S.}\ \bibnamefont {Barker}}, \bibinfo
  {author} {\bibfnamefont {D.}~\bibnamefont {Westly}}, \bibinfo {author}
  {\bibfnamefont {A.}~\bibnamefont {Yulaev}}, \bibinfo {author} {\bibfnamefont
  {N.}~\bibnamefont {Klimov}}, \bibinfo {author} {\bibfnamefont
  {A.}~\bibnamefont {Agrawal}}, \bibinfo {author} {\bibfnamefont
  {S.}~\bibnamefont {Eckel}}, \bibinfo {author} {\bibfnamefont
  {V.}~\bibnamefont {Aksyuk}},\ and\ \bibinfo {author} {\bibfnamefont {J.~J.}\
  \bibnamefont {McClelland}},\ }\bibfield  {title} {\bibinfo {title}
  {Magneto-optical trapping using planar optics},\ }\href
  {https://doi.org/10.1088/1367-2630/abdce3} {\bibfield  {journal} {\bibinfo
  {journal} {New Journal of Physics}\ }\textbf {\bibinfo {volume} {23}},\
  \bibinfo {pages} {013021} (\bibinfo {year} {2021})}\BibitemShut {NoStop}%
\bibitem [{\citenamefont {Chauhan}\ \emph {et~al.}(2019)\citenamefont
  {Chauhan}, \citenamefont {Bose}, \citenamefont {Puckett}, \citenamefont
  {Moreira}, \citenamefont {Nelson},\ and\ \citenamefont
  {Blumenthal}}]{Chauhan:19}%
  \BibitemOpen
  \bibfield  {author} {\bibinfo {author} {\bibfnamefont {N.}~\bibnamefont
  {Chauhan}}, \bibinfo {author} {\bibfnamefont {D.}~\bibnamefont {Bose}},
  \bibinfo {author} {\bibfnamefont {M.}~\bibnamefont {Puckett}}, \bibinfo
  {author} {\bibfnamefont {R.}~\bibnamefont {Moreira}}, \bibinfo {author}
  {\bibfnamefont {K.}~\bibnamefont {Nelson}},\ and\ \bibinfo {author}
  {\bibfnamefont {D.~J.}\ \bibnamefont {Blumenthal}},\ }\bibfield  {title}
  {\bibinfo {title} {Photonic integrated $\rm{Si}_3\rm{N}_4$ ultra-large-area
  grating waveguide mot interface for 3d atomic clock laser cooling},\ }in\
  \href {https://doi.org/10.1364/CLEO_SI.2019.STu4O.3} {\emph {\bibinfo
  {booktitle} {Conference on Lasers and Electro-Optics (2019)}}}\ (\bibinfo
  {publisher} {Optical Society of America},\ \bibinfo {year}
  {2019})\BibitemShut {NoStop}%
\bibitem [{\citenamefont {Isichenko}\ \emph {et~al.}(2022)\citenamefont
  {Isichenko}, \citenamefont {Chauhan}, \citenamefont {Bose}, \citenamefont
  {Kunz},\ and\ \citenamefont {Blumenthal}}]{Chauhan:22}%
  \BibitemOpen
  \bibfield  {author} {\bibinfo {author} {\bibfnamefont {A.}~\bibnamefont
  {Isichenko}}, \bibinfo {author} {\bibfnamefont {N.}~\bibnamefont {Chauhan}},
  \bibinfo {author} {\bibfnamefont {D.}~\bibnamefont {Bose}}, \bibinfo {author}
  {\bibfnamefont {P.~D.}\ \bibnamefont {Kunz}},\ and\ \bibinfo {author}
  {\bibfnamefont {D.~J.}\ \bibnamefont {Blumenthal}},\ }\bibfield  {title}
  {\bibinfo {title} {Cooling rubidium atoms with a photonic integrated 3d
  magneto-optical trap},\ }in\ \href
  {https://ocpi.ece.ucsb.edu/sites/default/files/2022-08/7D9CBE28-DEEC-45B9-B5BB01E4CC62D89B_so3783331_0.pdf}
  {\emph {\bibinfo {booktitle} {Optical Sensors and Sensing Congress (2022)}}}\
  (\bibinfo  {publisher} {Optical Society of America},\ \bibinfo {year}
  {2022})\BibitemShut {NoStop}%
\bibitem [{\citenamefont {Yulaev}\ \emph {et~al.}(2021)\citenamefont {Yulaev},
  \citenamefont {Zhu}, \citenamefont {Ropp}, \citenamefont {Westly},
  \citenamefont {Simelgor}, \citenamefont {Zhang}, \citenamefont {Lezec},
  \citenamefont {Agrawal},\ and\ \citenamefont {Aksyuk}}]{Yulaev:21}%
  \BibitemOpen
  \bibfield  {author} {\bibinfo {author} {\bibfnamefont {A.}~\bibnamefont
  {Yulaev}}, \bibinfo {author} {\bibfnamefont {W.}~\bibnamefont {Zhu}},
  \bibinfo {author} {\bibfnamefont {C.}~\bibnamefont {Ropp}}, \bibinfo {author}
  {\bibfnamefont {D.~A.}\ \bibnamefont {Westly}}, \bibinfo {author}
  {\bibfnamefont {G.}~\bibnamefont {Simelgor}}, \bibinfo {author}
  {\bibfnamefont {C.}~\bibnamefont {Zhang}}, \bibinfo {author} {\bibfnamefont
  {H.~J.}\ \bibnamefont {Lezec}}, \bibinfo {author} {\bibfnamefont
  {A.}~\bibnamefont {Agrawal}},\ and\ \bibinfo {author} {\bibfnamefont {V.~A.}\
  \bibnamefont {Aksyuk}},\ }\bibfield  {title} {\bibinfo {title} {Interfacing
  photonics to free-space via large-area inverse-designed diffraction elements
  and metasurfaces},\ }in\ \href {https://doi.org/10.1364/OFC.2021.F2B.1}
  {\emph {\bibinfo {booktitle} {Optical Fiber Communication Conference (OFC)
  2021}}}\ (\bibinfo  {publisher} {Optica Publishing Group},\ \bibinfo {year}
  {2021})\ p.\ \bibinfo {pages} {F2B.1}\BibitemShut {NoStop}%
\bibitem [{\citenamefont {Zhu}\ \emph {et~al.}(2020)\citenamefont {Zhu},
  \citenamefont {Liu}, \citenamefont {Sain}, \citenamefont {Wang},
  \citenamefont {Schlickriede}, \citenamefont {Tang}, \citenamefont {Deng},
  \citenamefont {Li}, \citenamefont {Yang}, \citenamefont {Holynski},
  \citenamefont {Zhang}, \citenamefont {Zentgraf}, \citenamefont {Bongs},
  \citenamefont {Lien},\ and\ \citenamefont {Li}}]{metasurface}%
  \BibitemOpen
  \bibfield  {author} {\bibinfo {author} {\bibfnamefont {L.}~\bibnamefont
  {Zhu}}, \bibinfo {author} {\bibfnamefont {X.}~\bibnamefont {Liu}}, \bibinfo
  {author} {\bibfnamefont {B.}~\bibnamefont {Sain}}, \bibinfo {author}
  {\bibfnamefont {M.}~\bibnamefont {Wang}}, \bibinfo {author} {\bibfnamefont
  {C.}~\bibnamefont {Schlickriede}}, \bibinfo {author} {\bibfnamefont
  {Y.}~\bibnamefont {Tang}}, \bibinfo {author} {\bibfnamefont {J.}~\bibnamefont
  {Deng}}, \bibinfo {author} {\bibfnamefont {K.}~\bibnamefont {Li}}, \bibinfo
  {author} {\bibfnamefont {J.}~\bibnamefont {Yang}}, \bibinfo {author}
  {\bibfnamefont {M.}~\bibnamefont {Holynski}}, \bibinfo {author}
  {\bibfnamefont {S.}~\bibnamefont {Zhang}}, \bibinfo {author} {\bibfnamefont
  {T.}~\bibnamefont {Zentgraf}}, \bibinfo {author} {\bibfnamefont
  {K.}~\bibnamefont {Bongs}}, \bibinfo {author} {\bibfnamefont {Y.-H.}\
  \bibnamefont {Lien}},\ and\ \bibinfo {author} {\bibfnamefont
  {G.}~\bibnamefont {Li}},\ }\bibfield  {title} {\bibinfo {title} {A dielectric
  metasurface optical chip for the generation of cold atoms},\ }\href
  {https://doi.org/10.1126/sciadv.abb6667} {\bibfield  {journal} {\bibinfo
  {journal} {Science Advances}\ }\textbf {\bibinfo {volume} {6}},\ \bibinfo
  {pages} {eabb6667} (\bibinfo {year} {2020})}\BibitemShut {NoStop}%
\bibitem [{\citenamefont {Nshii}\ \emph {et~al.}(2013)\citenamefont {Nshii},
  \citenamefont {Vangeleyn}, \citenamefont {Cotter}, \citenamefont {Griffin},
  \citenamefont {Hinds}, \citenamefont {Ironside}, \citenamefont {See},
  \citenamefont {Sinclair}, \citenamefont {Riis},\ and\ \citenamefont
  {Arnold}}]{Nshii2013}%
  \BibitemOpen
  \bibfield  {author} {\bibinfo {author} {\bibfnamefont {C.~C.}\ \bibnamefont
  {Nshii}}, \bibinfo {author} {\bibfnamefont {M.}~\bibnamefont {Vangeleyn}},
  \bibinfo {author} {\bibfnamefont {J.~P.}\ \bibnamefont {Cotter}}, \bibinfo
  {author} {\bibfnamefont {P.~F.}\ \bibnamefont {Griffin}}, \bibinfo {author}
  {\bibfnamefont {E.~A.}\ \bibnamefont {Hinds}}, \bibinfo {author}
  {\bibfnamefont {C.~N.}\ \bibnamefont {Ironside}}, \bibinfo {author}
  {\bibfnamefont {P.}~\bibnamefont {See}}, \bibinfo {author} {\bibfnamefont
  {A.~G.}\ \bibnamefont {Sinclair}}, \bibinfo {author} {\bibfnamefont
  {E.}~\bibnamefont {Riis}},\ and\ \bibinfo {author} {\bibfnamefont {A.~S.}\
  \bibnamefont {Arnold}},\ }\bibfield  {title} {\bibinfo {title} {A
  surface-patterned chip as a strong source of ultracold atoms for quantum
  technologies},\ }\href {https://doi.org/10.1038/nnano.2013.47} {\bibfield
  {journal} {\bibinfo  {journal} {Nature Nanotechnology}\ }\textbf {\bibinfo
  {volume} {8}},\ \bibinfo {pages} {321} (\bibinfo {year} {2013})}\BibitemShut
  {NoStop}%
\bibitem [{\citenamefont {Henderson}\ \emph {et~al.}(2020)\citenamefont
  {Henderson}, \citenamefont {Johnson}, \citenamefont {Kale}, \citenamefont
  {Griffin}, \citenamefont {Riis},\ and\ \citenamefont
  {Arnold}}]{Henderson:20}%
  \BibitemOpen
  \bibfield  {author} {\bibinfo {author} {\bibfnamefont {V.~A.}\ \bibnamefont
  {Henderson}}, \bibinfo {author} {\bibfnamefont {M.~Y.~H.}\ \bibnamefont
  {Johnson}}, \bibinfo {author} {\bibfnamefont {Y.~B.}\ \bibnamefont {Kale}},
  \bibinfo {author} {\bibfnamefont {P.~F.}\ \bibnamefont {Griffin}}, \bibinfo
  {author} {\bibfnamefont {E.}~\bibnamefont {Riis}},\ and\ \bibinfo {author}
  {\bibfnamefont {A.~S.}\ \bibnamefont {Arnold}},\ }\bibfield  {title}
  {\bibinfo {title} {Optical characterisation of micro-fabricated fresnel zone
  plates for atomic waveguides},\ }\href {https://doi.org/10.1364/OE.388897}
  {\bibfield  {journal} {\bibinfo  {journal} {Opt. Express}\ }\textbf {\bibinfo
  {volume} {28}},\ \bibinfo {pages} {9072} (\bibinfo {year}
  {2020})}\BibitemShut {NoStop}%
\bibitem [{\citenamefont {McGilligan}\ \emph {et~al.}(2017)\citenamefont
  {McGilligan}, \citenamefont {Griffin}, \citenamefont {Elvin}, \citenamefont
  {Ingleby}, \citenamefont {Riis},\ and\ \citenamefont
  {Arnold}}]{McGilligan2017}%
  \BibitemOpen
  \bibfield  {author} {\bibinfo {author} {\bibfnamefont {J.~P.}\ \bibnamefont
  {McGilligan}}, \bibinfo {author} {\bibfnamefont {P.~F.}\ \bibnamefont
  {Griffin}}, \bibinfo {author} {\bibfnamefont {R.}~\bibnamefont {Elvin}},
  \bibinfo {author} {\bibfnamefont {S.~J.}\ \bibnamefont {Ingleby}}, \bibinfo
  {author} {\bibfnamefont {E.}~\bibnamefont {Riis}},\ and\ \bibinfo {author}
  {\bibfnamefont {A.~S.}\ \bibnamefont {Arnold}},\ }\bibfield  {title}
  {\bibinfo {title} {Grating chips for quantum technologies},\ }\href
  {https://doi.org/10.1038/s41598-017-00254-0} {\bibfield  {journal} {\bibinfo
  {journal} {Scientific Reports}\ }\textbf {\bibinfo {volume} {7}},\ \bibinfo
  {pages} {384} (\bibinfo {year} {2017})}\BibitemShut {NoStop}%
\bibitem [{\citenamefont {Chen}\ \emph {et~al.}(2022)\citenamefont {Chen},
  \citenamefont {Huang}, \citenamefont {Xu}, \citenamefont {Zhang},
  \citenamefont {Ma}, \citenamefont {Lu}, \citenamefont {Wang}, \citenamefont
  {Chen}, \citenamefont {Zhang}, \citenamefont {Tang}, \citenamefont {Dong},
  \citenamefont {Liu}, \citenamefont {Xiang}, \citenamefont {Guo},\ and\
  \citenamefont {Zou}}]{chen2021planar}%
  \BibitemOpen
  \bibfield  {author} {\bibinfo {author} {\bibfnamefont {L.}~\bibnamefont
  {Chen}}, \bibinfo {author} {\bibfnamefont {C.-J.}\ \bibnamefont {Huang}},
  \bibinfo {author} {\bibfnamefont {X.-B.}\ \bibnamefont {Xu}}, \bibinfo
  {author} {\bibfnamefont {Y.-C.}\ \bibnamefont {Zhang}}, \bibinfo {author}
  {\bibfnamefont {D.-Q.}\ \bibnamefont {Ma}}, \bibinfo {author} {\bibfnamefont
  {Z.-T.}\ \bibnamefont {Lu}}, \bibinfo {author} {\bibfnamefont {Z.-B.}\
  \bibnamefont {Wang}}, \bibinfo {author} {\bibfnamefont {G.-J.}\ \bibnamefont
  {Chen}}, \bibinfo {author} {\bibfnamefont {J.-Z.}\ \bibnamefont {Zhang}},
  \bibinfo {author} {\bibfnamefont {H.~X.}\ \bibnamefont {Tang}}, \bibinfo
  {author} {\bibfnamefont {C.-H.}\ \bibnamefont {Dong}}, \bibinfo {author}
  {\bibfnamefont {W.}~\bibnamefont {Liu}}, \bibinfo {author} {\bibfnamefont
  {G.-Y.}\ \bibnamefont {Xiang}}, \bibinfo {author} {\bibfnamefont {G.-C.}\
  \bibnamefont {Guo}},\ and\ \bibinfo {author} {\bibfnamefont {C.-L.}\
  \bibnamefont {Zou}},\ }\bibfield  {title} {\bibinfo {title}
  {Planar-integrated magneto-optical trap},\ }\href
  {https://doi.org/10.1103/PhysRevApplied.17.034031} {\bibfield  {journal}
  {\bibinfo  {journal} {Phys. Rev. Applied}\ }\textbf {\bibinfo {volume}
  {17}},\ \bibinfo {pages} {034031} (\bibinfo {year} {2022})}\BibitemShut
  {NoStop}%
\bibitem [{\citenamefont {Kang}\ \emph {et~al.}(2019)\citenamefont {Kang},
  \citenamefont {Moore}, \citenamefont {McGilligan}, \citenamefont {Mott},
  \citenamefont {Mis}, \citenamefont {Roper}, \citenamefont {Donley},\ and\
  \citenamefont {Kitching}}]{Kang:19}%
  \BibitemOpen
  \bibfield  {author} {\bibinfo {author} {\bibfnamefont {S.}~\bibnamefont
  {Kang}}, \bibinfo {author} {\bibfnamefont {K.~R.}\ \bibnamefont {Moore}},
  \bibinfo {author} {\bibfnamefont {J.~P.}\ \bibnamefont {McGilligan}},
  \bibinfo {author} {\bibfnamefont {R.}~\bibnamefont {Mott}}, \bibinfo {author}
  {\bibfnamefont {A.}~\bibnamefont {Mis}}, \bibinfo {author} {\bibfnamefont
  {C.}~\bibnamefont {Roper}}, \bibinfo {author} {\bibfnamefont {E.~A.}\
  \bibnamefont {Donley}},\ and\ \bibinfo {author} {\bibfnamefont
  {J.}~\bibnamefont {Kitching}},\ }\bibfield  {title} {\bibinfo {title}
  {Magneto-optic trap using a reversible, solid-state alkali-metal source},\
  }\href {https://doi.org/10.1364/OL.44.003002} {\bibfield  {journal} {\bibinfo
   {journal} {Optics Letters}\ }\textbf {\bibinfo {volume} {44}},\ \bibinfo
  {pages} {3002} (\bibinfo {year} {2019})}\BibitemShut {NoStop}%
\bibitem [{\citenamefont {McGilligan}\ \emph
  {et~al.}(2020{\natexlab{a}})\citenamefont {McGilligan}, \citenamefont
  {Moore}, \citenamefont {Kang}, \citenamefont {Mott}, \citenamefont {Mis},
  \citenamefont {Roper}, \citenamefont {Donley},\ and\ \citenamefont
  {Kitching}}]{mcgilliganAIB}%
  \BibitemOpen
  \bibfield  {author} {\bibinfo {author} {\bibfnamefont {J.~P.}\ \bibnamefont
  {McGilligan}}, \bibinfo {author} {\bibfnamefont {K.~R.}\ \bibnamefont
  {Moore}}, \bibinfo {author} {\bibfnamefont {S.}~\bibnamefont {Kang}},
  \bibinfo {author} {\bibfnamefont {R.}~\bibnamefont {Mott}}, \bibinfo {author}
  {\bibfnamefont {A.}~\bibnamefont {Mis}}, \bibinfo {author} {\bibfnamefont
  {C.}~\bibnamefont {Roper}}, \bibinfo {author} {\bibfnamefont {E.~A.}\
  \bibnamefont {Donley}},\ and\ \bibinfo {author} {\bibfnamefont
  {J.}~\bibnamefont {Kitching}},\ }\bibfield  {title} {\bibinfo {title}
  {Dynamic characterization of an alkali-ion battery as a source for
  laser-cooled atoms},\ }\href
  {https://doi.org/10.1103/PhysRevApplied.13.044038} {\bibfield  {journal}
  {\bibinfo  {journal} {Phys. Rev. Applied}\ }\textbf {\bibinfo {volume}
  {13}},\ \bibinfo {pages} {044038} (\bibinfo {year}
  {2020}{\natexlab{a}})}\BibitemShut {NoStop}%
\bibitem [{\citenamefont {McGilligan}\ \emph
  {et~al.}(2020{\natexlab{b}})\citenamefont {McGilligan}, \citenamefont
  {Moore}, \citenamefont {Dellis}, \citenamefont {Martinez}, \citenamefont
  {de~Clercq}, \citenamefont {Griffin}, \citenamefont {Arnold}, \citenamefont
  {Riis}, \citenamefont {Boudot},\ and\ \citenamefont
  {Kitching}}]{McGilligan2020}%
  \BibitemOpen
  \bibfield  {author} {\bibinfo {author} {\bibfnamefont {J.~P.}\ \bibnamefont
  {McGilligan}}, \bibinfo {author} {\bibfnamefont {K.~R.}\ \bibnamefont
  {Moore}}, \bibinfo {author} {\bibfnamefont {A.}~\bibnamefont {Dellis}},
  \bibinfo {author} {\bibfnamefont {G.~D.}\ \bibnamefont {Martinez}}, \bibinfo
  {author} {\bibfnamefont {E.}~\bibnamefont {de~Clercq}}, \bibinfo {author}
  {\bibfnamefont {P.~F.}\ \bibnamefont {Griffin}}, \bibinfo {author}
  {\bibfnamefont {A.~S.}\ \bibnamefont {Arnold}}, \bibinfo {author}
  {\bibfnamefont {E.}~\bibnamefont {Riis}}, \bibinfo {author} {\bibfnamefont
  {R.}~\bibnamefont {Boudot}},\ and\ \bibinfo {author} {\bibfnamefont
  {J.}~\bibnamefont {Kitching}},\ }\bibfield  {title} {\bibinfo {title} {Laser
  cooling in a chip-scale platform},\ }\href
  {https://doi.org/10.1063/5.0014658} {\bibfield  {journal} {\bibinfo
  {journal} {Applied Physics Letters}\ }\textbf {\bibinfo {volume} {117}},\
  \bibinfo {pages} {054001} (\bibinfo {year} {2020}{\natexlab{b}})}\BibitemShut
  {NoStop}%
\bibitem [{\citenamefont {Strangfeld}\ \emph {et~al.}(2022)\citenamefont
  {Strangfeld}, \citenamefont {Wiegand}, \citenamefont {Kluge}, \citenamefont
  {Schoch},\ and\ \citenamefont {Krutzik}}]{plugplay}%
  \BibitemOpen
  \bibfield  {author} {\bibinfo {author} {\bibfnamefont {A.}~\bibnamefont
  {Strangfeld}}, \bibinfo {author} {\bibfnamefont {B.}~\bibnamefont {Wiegand}},
  \bibinfo {author} {\bibfnamefont {J.}~\bibnamefont {Kluge}}, \bibinfo
  {author} {\bibfnamefont {M.}~\bibnamefont {Schoch}},\ and\ \bibinfo {author}
  {\bibfnamefont {M.}~\bibnamefont {Krutzik}},\ }\bibfield  {title} {\bibinfo
  {title} {{Compact plug and play optical frequency reference device based on
  Doppler-free spectroscopy of rubidium vapor}},\ }\href
  {https://doi.org/10.1364/OE.453942} {\bibfield  {journal} {\bibinfo
  {journal} {Opt. Express}\ }\textbf {\bibinfo {volume} {30}},\ \bibinfo
  {pages} {12039} (\bibinfo {year} {2022})}\BibitemShut {NoStop}%
\bibitem [{\citenamefont {Madkhaly}\ \emph {et~al.}(2021)\citenamefont
  {Madkhaly}, \citenamefont {Coles}, \citenamefont {Morley}, \citenamefont
  {Colquhoun}, \citenamefont {Fromhold}, \citenamefont {Cooper},\ and\
  \citenamefont {Hackerm\"uller}}]{3Dprint}%
  \BibitemOpen
  \bibfield  {author} {\bibinfo {author} {\bibfnamefont {S.}~\bibnamefont
  {Madkhaly}}, \bibinfo {author} {\bibfnamefont {L.}~\bibnamefont {Coles}},
  \bibinfo {author} {\bibfnamefont {C.}~\bibnamefont {Morley}}, \bibinfo
  {author} {\bibfnamefont {C.}~\bibnamefont {Colquhoun}}, \bibinfo {author}
  {\bibfnamefont {T.}~\bibnamefont {Fromhold}}, \bibinfo {author}
  {\bibfnamefont {N.}~\bibnamefont {Cooper}},\ and\ \bibinfo {author}
  {\bibfnamefont {L.}~\bibnamefont {Hackerm\"uller}},\ }\bibfield  {title}
  {\bibinfo {title} {Performance-optimized components for quantum technologies
  via additive manufacturing},\ }\href
  {https://doi.org/10.1103/PRXQuantum.2.030326} {\bibfield  {journal} {\bibinfo
   {journal} {PRX Quantum}\ }\textbf {\bibinfo {volume} {2}},\ \bibinfo {pages}
  {030326} (\bibinfo {year} {2021})}\BibitemShut {NoStop}%
\bibitem [{\citenamefont {Middlemiss}\ \emph {et~al.}(2016)\citenamefont
  {Middlemiss}, \citenamefont {Samarelli}, \citenamefont {Paul}, \citenamefont
  {Hough}, \citenamefont {Rowan},\ and\ \citenamefont
  {Hammond}}]{Middlemiss2016}%
  \BibitemOpen
  \bibfield  {author} {\bibinfo {author} {\bibfnamefont {R.~P.}\ \bibnamefont
  {Middlemiss}}, \bibinfo {author} {\bibfnamefont {A.}~\bibnamefont
  {Samarelli}}, \bibinfo {author} {\bibfnamefont {D.~J.}\ \bibnamefont {Paul}},
  \bibinfo {author} {\bibfnamefont {J.}~\bibnamefont {Hough}}, \bibinfo
  {author} {\bibfnamefont {S.}~\bibnamefont {Rowan}},\ and\ \bibinfo {author}
  {\bibfnamefont {G.~D.}\ \bibnamefont {Hammond}},\ }\bibfield  {title}
  {\bibinfo {title} {Measurement of the earth tides with a mems gravimeter},\
  }\href {https://doi.org/10.1038/nature17397} {\bibfield  {journal} {\bibinfo
  {journal} {Nature}\ }\textbf {\bibinfo {volume} {531}},\ \bibinfo {pages}
  {614} (\bibinfo {year} {2016})}\BibitemShut {NoStop}%
\bibitem [{\citenamefont {Griffin}\ \emph {et~al.}(2005)\citenamefont
  {Griffin}, \citenamefont {Weatherill},\ and\ \citenamefont
  {Adams}}]{GriffLaser}%
  \BibitemOpen
  \bibfield  {author} {\bibinfo {author} {\bibfnamefont {P.~F.}\ \bibnamefont
  {Griffin}}, \bibinfo {author} {\bibfnamefont {K.~J.}\ \bibnamefont
  {Weatherill}},\ and\ \bibinfo {author} {\bibfnamefont {C.~S.}\ \bibnamefont
  {Adams}},\ }\bibfield  {title} {\bibinfo {title} {{Fast switching of alkali
  atom dispensers using laser-induced heating}},\ }\href
  {https://doi.org/10.1063/1.2038167} {\bibfield  {journal} {\bibinfo
  {journal} {Rev. Sci. Instrum.}\ }\textbf {\bibinfo {volume} {76}},\ \bibinfo
  {pages} {93102} (\bibinfo {year} {2005})}\BibitemShut {NoStop}%
\bibitem [{\citenamefont {Dyer}\ \emph {et~al.}(2022)\citenamefont {Dyer},
  \citenamefont {Griffin}, \citenamefont {Arnold}, \citenamefont {Mirando},
  \citenamefont {Burt}, \citenamefont {Riis},\ and\ \citenamefont
  {McGilligan}}]{dyer}%
  \BibitemOpen
  \bibfield  {author} {\bibinfo {author} {\bibfnamefont {S.}~\bibnamefont
  {Dyer}}, \bibinfo {author} {\bibfnamefont {P.~F.}\ \bibnamefont {Griffin}},
  \bibinfo {author} {\bibfnamefont {A.~S.}\ \bibnamefont {Arnold}}, \bibinfo
  {author} {\bibfnamefont {F.}~\bibnamefont {Mirando}}, \bibinfo {author}
  {\bibfnamefont {D.~P.}\ \bibnamefont {Burt}}, \bibinfo {author}
  {\bibfnamefont {E.}~\bibnamefont {Riis}},\ and\ \bibinfo {author}
  {\bibfnamefont {J.~P.}\ \bibnamefont {McGilligan}},\ }\bibfield  {title}
  {\bibinfo {title} {Micro-machined deep silicon atomic vapor cells},\ }\href
  {https://doi.org/10.1063/5.0114762} {\bibfield  {journal} {\bibinfo
  {journal} {Journal of Applied Physics}\ }\textbf {\bibinfo {volume} {132}},\
  \bibinfo {pages} {134401} (\bibinfo {year} {2022})}\BibitemShut {NoStop}%
\bibitem [{\citenamefont {Bregazzi}\ \emph {et~al.}(2021)\citenamefont
  {Bregazzi}, \citenamefont {Griffin}, \citenamefont {Arnold}, \citenamefont
  {Burt}, \citenamefont {Martinez}, \citenamefont {Boudot}, \citenamefont
  {Kitching}, \citenamefont {Riis},\ and\ \citenamefont
  {McGilligan}}]{Bregazzi}%
  \BibitemOpen
  \bibfield  {author} {\bibinfo {author} {\bibfnamefont {A.}~\bibnamefont
  {Bregazzi}}, \bibinfo {author} {\bibfnamefont {P.~F.}\ \bibnamefont
  {Griffin}}, \bibinfo {author} {\bibfnamefont {A.~S.}\ \bibnamefont {Arnold}},
  \bibinfo {author} {\bibfnamefont {D.~P.}\ \bibnamefont {Burt}}, \bibinfo
  {author} {\bibfnamefont {G.}~\bibnamefont {Martinez}}, \bibinfo {author}
  {\bibfnamefont {R.}~\bibnamefont {Boudot}}, \bibinfo {author} {\bibfnamefont
  {J.}~\bibnamefont {Kitching}}, \bibinfo {author} {\bibfnamefont
  {E.}~\bibnamefont {Riis}},\ and\ \bibinfo {author} {\bibfnamefont {J.~P.}\
  \bibnamefont {McGilligan}},\ }\bibfield  {title} {\bibinfo {title} {A simple
  imaging solution for chip-scale laser cooling},\ }\href
  {https://doi.org/10.1063/5.0068725} {\bibfield  {journal} {\bibinfo
  {journal} {Applied Physics Letters}\ }\textbf {\bibinfo {volume} {119}},\
  \bibinfo {pages} {184002} (\bibinfo {year} {2021})}\BibitemShut {NoStop}%
\bibitem [{\citenamefont {Lecomte}\ \emph {et~al.}(2000)\citenamefont
  {Lecomte}, \citenamefont {Fretel}, \citenamefont {Mileti},\ and\
  \citenamefont {Thomann}}]{GaetanoZeeman}%
  \BibitemOpen
  \bibfield  {author} {\bibinfo {author} {\bibfnamefont {S.}~\bibnamefont
  {Lecomte}}, \bibinfo {author} {\bibfnamefont {E.}~\bibnamefont {Fretel}},
  \bibinfo {author} {\bibfnamefont {G.}~\bibnamefont {Mileti}},\ and\ \bibinfo
  {author} {\bibfnamefont {P.}~\bibnamefont {Thomann}},\ }\bibfield  {title}
  {\bibinfo {title} {Self-aligned extended-cavity diode laser stabilized by the
  zeeman effect on the cesium d2 line},\ }\href
  {https://doi.org/10.1364/AO.39.001426} {\bibfield  {journal} {\bibinfo
  {journal} {Appl. Opt.}\ }\textbf {\bibinfo {volume} {39}},\ \bibinfo {pages}
  {1426} (\bibinfo {year} {2000})}\BibitemShut {NoStop}%
\bibitem [{\citenamefont {Donley}\ \emph {et~al.}(2005)\citenamefont {Donley},
  \citenamefont {Heavner}, \citenamefont {Levi}, \citenamefont {Tataw},\ and\
  \citenamefont {Jefferts}}]{AOMtune}%
  \BibitemOpen
  \bibfield  {author} {\bibinfo {author} {\bibfnamefont {E.~A.}\ \bibnamefont
  {Donley}}, \bibinfo {author} {\bibfnamefont {T.~P.}\ \bibnamefont {Heavner}},
  \bibinfo {author} {\bibfnamefont {F.}~\bibnamefont {Levi}}, \bibinfo {author}
  {\bibfnamefont {M.~O.}\ \bibnamefont {Tataw}},\ and\ \bibinfo {author}
  {\bibfnamefont {S.~R.}\ \bibnamefont {Jefferts}},\ }\bibfield  {title}
  {\bibinfo {title} {{Double-pass acousto-optic modulator system}},\ }\href
  {https://doi.org/10.1063/1.1930095} {\bibfield  {journal} {\bibinfo
  {journal} {Rev. Sci. Instrum.}\ }\textbf {\bibinfo {volume} {76}},\ \bibinfo
  {pages} {63112} (\bibinfo {year} {2005})}\BibitemShut {NoStop}%
\bibitem [{\citenamefont {Steck}(2001)}]{Steck}%
  \BibitemOpen
  \bibfield  {author} {\bibinfo {author} {\bibfnamefont {D.~A.}\ \bibnamefont
  {Steck}},\ }\href {http://steck.us/alkalidata} {\bibinfo {title} {Rubidium 87
  d line data}},\ \bibinfo {howpublished} {Available Online} (\bibinfo {year}
  {2001})\BibitemShut {NoStop}%
\end{thebibliography}%

\end{document}